\documentclass[11pt,twoside]{article}

\usepackage{galev06}
\usepackage{epsf}
\usepackage{psfig}
\usepackage{lscape}

\begin{document}
\title{Biased galaxy formation in the fields of high-redshift AGN}   
\author{J.\ A.\ Stevens,$^1$ M.\ J.\ Page,$^2$ F.\ J.\ Carrera,$^3$ R.\ J.\
  Ivison,$^4$ and Ian Smail$^5$}   
\affil{$^1$Centre for Astrophysics Research, University of Hertfordshire,
  College Lane, Herts AL10 9AB, UK\\
$^2$Mullard Space Science Laboratory, University College London, Holmbury
  St. Mary, Dorking, Surrey RH5 6NT, UK\\
$^3$Instituto de Fisica de Cantabria (CSIC-UC), Avienida de los Castros 39005
  Santander, Spain\\
$^4$UK Astronomy Technology Centre, Royal Observatory, Blackford Hill,
  Edinburgh EH9 3HJ, UK\\
$^5$Institute for Computational Cosmology, Durham University, South Road,
  Durham DH1 3LE, UK 
}    

\begin{abstract} 
We discuss preliminary results from our programme to map the fields of
high-redshift AGN. In the context of the hierarchical models such fields are
predicted to contain an over-density of young, luminous galaxies destined to
evolve into the core of a rich cluster by the present epoch.  We have thus
imaged from submillimetre to X-ray wavelengths the few-arcmin scale fields of a
small sample of high-redshift QSOs. We
find that submillimetre wavelength data from SCUBA show striking over-densities
of luminous star-forming galaxies over scales of $\sim500$~kpc. Whilst many of
these galaxies are undetected even in deep near-IR imaging almost all of them
are detected by {\em Spitzer\/} at 4.5, 8.0 and 24~$\mu$m, showing that they
have extremely red colours. However, they are not detected in our {\em
XMM-Newton\/} observations suggesting that any AGN must be highly
obscured. Optical-through-mid-IR SEDs show the redshifted 1.6~$\mu$m bump from
star-light giving preliminary evidence that the galaxies lie at the same
redshift, and thus in the same structure, as the QSO although this finding must
be confirmed with photometric and/or spectroscopic redshifts.
\end{abstract}



\section{Introduction}

The popular hierarchical model of galaxy formation predicts that elliptical
galaxies found today in the cores of rich clusters formed at high redshifts and
at rare high-density peaks of the dark matter distribution. Within these
regions, gas rich proto-galaxies merge together rapidly and form stars at a
high rate. The same reservoir of gas used to build the stellar mass can also
fuel the growth of the supermassive black holes (SMBH) found dormant in the
centre of the galaxies at low redshift \citep{kah00}. Since massive star
formation is known to be a dusty phenomenon, the light from such young galaxies
may well be highly obscured rendering the population inaccessible to optical
techniques. Dust enshrouded star formation is, however, a luminous phenomenon
in the submillimetre (hereafter submm) through far-IR waveband where the
star-light absorbed by dust grains is re-emitted.

Unfortunately, it is not currently feasible at submm/far-IR wavelengths to
perform a survey of a randomly selected region of sky over an area sufficiently
large to ensure that rare structures in the early universe are contained within
it. However, one method of locating such structures is to target the fields of
high-redshift AGN. Given their huge luminosities, such objects must already
contain a SMBH, and therefore must represent some of the most massive objects
in existence at their epochs. They should thus act as signposts to the rare
high-density peaks that we wish to study and can target with existing
technology at submm wavelengths \citep{ivi00,ste03}. This paper presents preliminary
results from such a programme.  We concentrate on two QSO fields at $1<z<3$,
presenting submm results obtained with SCUBA on the JCMT and follow-up imaging
at near-IR (UKIRT) and mid-IR ({\em Spitzer}) wavelengths.

\section{Over-densities of luminous, extremely red star-forming galaxies}

The two targets discussed here are the X-ray absorbed QSOs
RX\,J$094144.51+385434.8$ ($z=1.82$) and RX\,J$121803.82+470854.6$
($z=1.74$). The importance of these objects has been discussed extensively in
the literature \citep{pag04,ste05}. See also the contribution by
\citeauthor{pag07} in this volume.

\begin{figure}[!ht]
\plotone{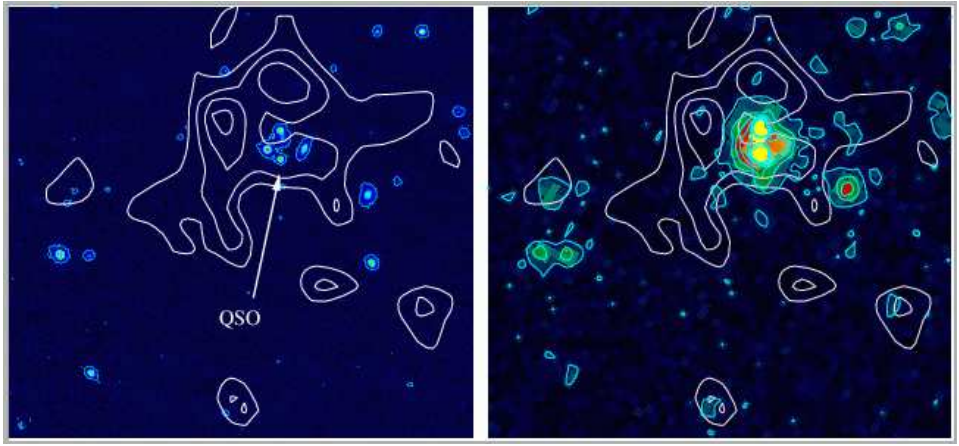}
\caption{Submm, mid- and near-IR imaging of the
  RX\,J$121803.82+470854.6$ field. Left-hand panel shows 850~$\mu$m
  contours (2,3,4,5\ $\sigma$)
  on a deep $K$-band image whilst the right-hand panel shows the same
  contours on an 8.0~$\mu$m image. Note that dust peaks lacking
  $K$-band counterparts often do have detections at 8.0~$\mu$m. Each panel is
  $\sim 1$ arcmin square.}
\label{fig:fig1}
\end{figure}

Submm, near- and mid-IR images of the RX\,J$121803.82+470854.6$ field are shown
in Fig.~\ref{fig:fig1}. The 850~$\mu$m data reveal a large over-density of
sources, particularly pronounced in the close vicinity ($\sim 30$ arcsec) of
the QSO. Results from blank-field surveys predict, at most, one $\ge3 \ \sigma$
source in addition to the QSO. The near- and mid-IR
imaging gives an indication that many of these submm sources are indeed real
and have extremely red colours. Many of the them are detected by {\em
Spitzer\/} even when they are not seen in deep ($K \sim 20.5$) near-IR
imaging; e.g. the two most significant dust peaks in
Fig.~\ref{fig:fig1}. Star-formation rates computed in the standard manner are
at least several 100 M$_{\odot}$~yr$^{-1}$. It can thus be concluded that
we have discovered a large over-density of star-forming galaxies in the few
hundred kpc-scale environment of the QSO.

\begin{figure}[!ht]
\plotone{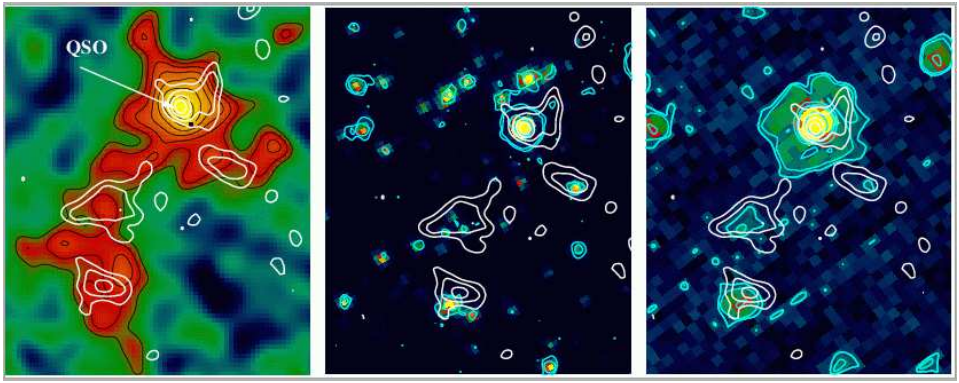}
\caption{Submm and mid-IR imaging of the
  RX\,J$094144.51+385434.8$ field. 450~$\mu$m contours
  on the 850~$\mu$m image (left), 450~$\mu$m contours
  on the 4.5~$\mu$m image (centre) and 450~$\mu$m
  contours on the 24~$\mu$m image (right). Submm contours start at
  2~$\sigma$ and increase in 1~$\sigma$ steps. Panels are $\sim1.0 \times 1.5$
  arcmin. The structure of galaxies is $\sim400$ kpc in extent \citep{ste04}.}
\label{fig:fig2}
\end{figure}

Fig.~\ref{fig:fig2} shows submm and mid-IR data for the
RX\,J$094144.51+385434.8$ field. Here we see a similar over-density of luminous
star-forming galaxies to that discussed above. For this dataset the observing
conditions were sufficiently transparent that we were able to collect good data
at 450~$\mu$m. These data show that the large-scale filamentary structure seen
at 850~$\mu$m breaks up into individual star-forming galaxies when observed at
higher resolution ($\sim8$ cf. $\sim14$ arcsec). Such datasets were very rarely
obtained with SCUBA at 450~$\mu$m because of the difficulty of observing at
short submm wavelengths from Mauna Kea. However, the increased sensitivity and
field-of-view afforded by SCUBA-2 will go a long way to alleviating this
problem and we can look forwards to higher resolution images of distant
galaxies.

\begin{figure}[!ht]
\plotfiddle{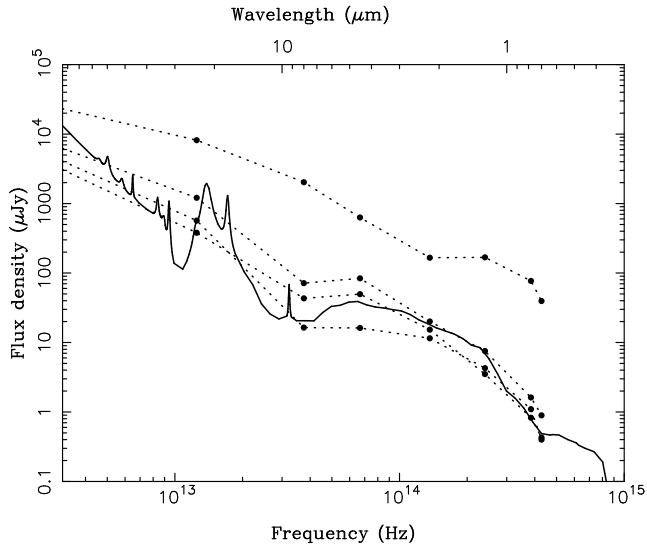}{2.6in}{270}{40}{40}{-170}{220}
\caption{Optical through mid-IR spectra of the QSO
  RX\,J$094144.51+385434.8$ (top dotted line and points) and the submm
  galaxies in its vicinity. The solid line is the spectrum of the local ULIRG
  Arp 220 redshifted to $z=1.82$ and scaled arbitrarily \citep{sil98}.}
\label{fig:fig3}
\end{figure}

A second result is that the submm-selected galaxies are detected by {\em
Spitzer\/} with both IRAC and MIPS (Fig.~\ref{fig:fig2}) verifying their
reality and, with the addition of optical and near-infrared imaging, allowing
us to construct SEDs of the individual galaxies (see Fig.~\ref{fig:fig3}). We
find that the SEDs are quite similar to that of the local ULIRG Arp~220 if it
is moved to the redshift of the QSO ($z=1.82$) suggesting that the bump we see
in the spectra, peaking roughly in the 4.5~$\mu$m band is the redshifted
1.6~$\mu$m bump from star-light. This result provides the first evidence that
the luminous star-forming galaxies discovered by SCUBA are at the same redshift
and in the same structure as the QSO. We are currently working to confirm this
finding with photometric and spectroscopic redshifts.

\subsection{Buried AGN?}

If these galaxies are to evolve into cluster elliptical galaxies then the SMBHs
found at their centres at the current epoch should be active at $z\sim2$. Our
50~ks {\em XMM-Newton\/} observations detect the QSOs but not the star-forming
galaxies in their fields. If we assume that these galaxies contain an AGN with
a luminosity $L(2-10\ {\rm keV})=10^{44}$ erg\ s$^{-1}$ and a photon index
$\Gamma=2$ then the non-detections imply a column density $N_H>5\times10^{23}$
cm$^{-2}$. Therefore, if these galaxies contain reasonably powerful AGN they
must be highly obscured.

\section{Looking ahead to SCUBA-2 and {\em Herschel}}

The next year will see major new facilities operating at far-IR and submm
wavelengths. Both {\em Herschel\/} and SCUBA-2 will provide much faster mapping
speeds than those currently available allowing us to map an area of sky around
the QSOs large enough to observe the whole filamentary structure of the
collapsing proto-cluster rather than just the core region. Simulations show
that this will require a map of at least $20$ Mpc square at $z=2$ (comoving
units). At present a significant amount of telescope time is required to map
one hundredth of this area but SCUBA-2 can map a region this large in only a
few hours.



\end{document}